\documentclass[aps,epl,twocolumn,groupedaddress,floatfix]{revtex4-1}

\usepackage{graphicx}
\usepackage{natbib}
\usepackage[colorlinks=false]{hyperref}
\usepackage{url}
\usepackage{enumerate}

\usepackage{amsmath,amssymb}
\usepackage{psfrag,epsfig}
\usepackage[utf8]{inputenc}

\newcommand{\rhoc}{\rho_{\scriptscriptstyle \rm c}}
\newcommand{\phiT}{\phi_{\scriptscriptstyle \rm T}}
\newcommand{\phiL}{\phi_{\scriptscriptstyle \rm L}}
\newcommand{\gT}{C_{\scriptscriptstyle \rm T}}
\newcommand{\gL}{C_{\scriptscriptstyle \rm L}}

\newcommand{\taurel}{\tau_{\scriptscriptstyle \rm rel}}
\newcommand{\kB}{k_{\scriptscriptstyle \rm B}}

\begin{document}
\setlength{\parskip}{0.3\baselineskip}

\title{Driving kinetically constrained models into non-equilibrium
  steady states: \\ Structural and slow transport properties}

\author{Francesco Turci}
\author{Estelle Pitard}
\affiliation{Laboratoire Charles Coulomb, Universit\'e Montpellier  II and CNRS, 34095 Montpellier, France}

\author{Mauro Sellitto} \affiliation{Department of Information
  Engineering , Second University of Naples, I-81031 Aversa (CE),
  Italy}



\begin{abstract}
Complex fluids in shear flow and biased dynamics in crowded
environments exhibit counterintuitive features which are difficult to
address both at a theoretical level and by molecular dynamic
simulations. To understand some of these features we study a schematic
model of a highly viscous liquid, the two-dimensional Kob-Andersen kinetically
constrained model, driven into non-equilibrium steady states by a
uniform non-Hamiltonian force.  We present a detailed numerical
analysis of the microscopic behavior of the model, including
transversal and longitudinal spatial correlations and dynamic
heterogeneities. In particular, we show that at high particle density
the transition from positive to negative resistance regimes in the
current vs field relation can be explained via the emergence of
nontrivial structures that intermittently trap the particles and slow
down the dynamics.  We relate such spatial structures to the current
vs field relation in the different transport regimes.
\end{abstract}

\pacs{}

\maketitle


\section{Introduction}

Slow relaxation and anomalous diffusion are common features of
disordered systems.  In particular, viscous liquids and highly packed
matter show dynamically arrested states (the glassy and the jammed
state respectively) characterized by steeply increasing relaxation
times and spatially heterogeneous and intermittent dynamics~\cite{DH,heterogeneities}.
Similar behavior has been also observed in sheared complex fluids and
dense granular materials~\cite{larson,fragile,jamming}, though in these
latter cases it is much more difficult to understand as it requires a
full dynamical description, due to the absence of a Boltzmann-Gibbs
framework.

The microscopic origin for these peculiar dynamically arrested states
has been the subject of many studies.  It has been shown that gels,
colloids, and supercooled liquids generally exhibit rare mobility
regions, and that an increase of mobility due to local relaxation
events facilitates the dynamics, allowing other regions to participate
cooperatively. Such facilitation mechanism has been supposed to be one
of the reasons underlying the slow dynamics and is at the origin of a vast
class of models called Kinetically Constrained Models (KCM), see
\cite{Ritort-Sollich,kcms} for reviews.

These models are very simple from a thermodynamic point of view as
they do not rely on any specific interaction potentials, but rather on
particular dynamic evolution rules.  They can be implemented under two
closely related forms: facilitated spin systems or kinetically
constrained lattice gases. In the first case, spins represent mobile
(or active) regions that flip between two or more states depending on
the status of the nearest neighbors, which can either facilitate or
forbid some spin flip. In the second case, the particle dynamics on a
lattice follows some specific kinetic rule facilitating or suppressing
some particle moves depending on the nearby local particle density. In
the spin case, the control parameter is the temperature defining the
density of excited states, while in the second case the particle
density (i.e the packing fraction) plays a central role. Both dynamic
evolution rules aim at simulating the cage effect due to the steric
hindrance among particles or spins belonging to the same dense region.

We consider here a specific case of the latter class of
models, the two-dimensional (2D) Kob-Andersen model with an externally applied field,
as introduced in Ref.~\cite{sellitto}.
%


%
  
In ref.~\cite{sellitto}, it has been shown by numerical simulation
that at high density, the model features a crossover from a flowing
(positive resistance) regime at a small field to a negative differential
resistance regime at a larger field.  The latter regime is accompanied
by unusual transport properties, including non-monotonic field
dependence of the structural relaxation time and rheological-like
behavior~\cite{sellitto}. Notably, the asymptotic large-deviation
limit in which the fluctuation relation holds is hardly attained on
the simulation timescale~\cite{sellitto2,TurciPitard}.  In
ref.~\cite{TurciPitard}, the anomalous space-time behavior of the
system has been quantified by providing a description in terms of a
field dependent dynamical transition between a flowing and blocked
phase with the use of the language of the \textit{thermodynamic of
  histories}.  This formalism has been used to evidence such a
dynamical phase transition in undriven glassy systems
\cite{histories}.  The present paper is an extended version of
Refs.~\cite{sellitto,TurciPitard} in which numerical simulation
results (some of which were previously announced only) are now fully
reported and are better understood in terms of a theoretical
approach. In particular, we give a microscopic description in
configuration space of the two transport regimes and describe the
nontrivial dynamical heterogeneities induced by the driving force. The
paper is organized as follows: in section \ref{model} we present the
description of the model, with the characterization of the
relationship between the current, the density of particles and the
driving field in section \ref{current}; in section \ref{sptm} we
discuss the role of heterogeneities; in section \ref{corr} we compute
global space correlation functions and in section \ref{traps} we
relate the microscopic structure (traps and domain walls) to the
current.

\begin{figure}[t]
\begin{center}
\includegraphics[width=0.5\columnwidth]{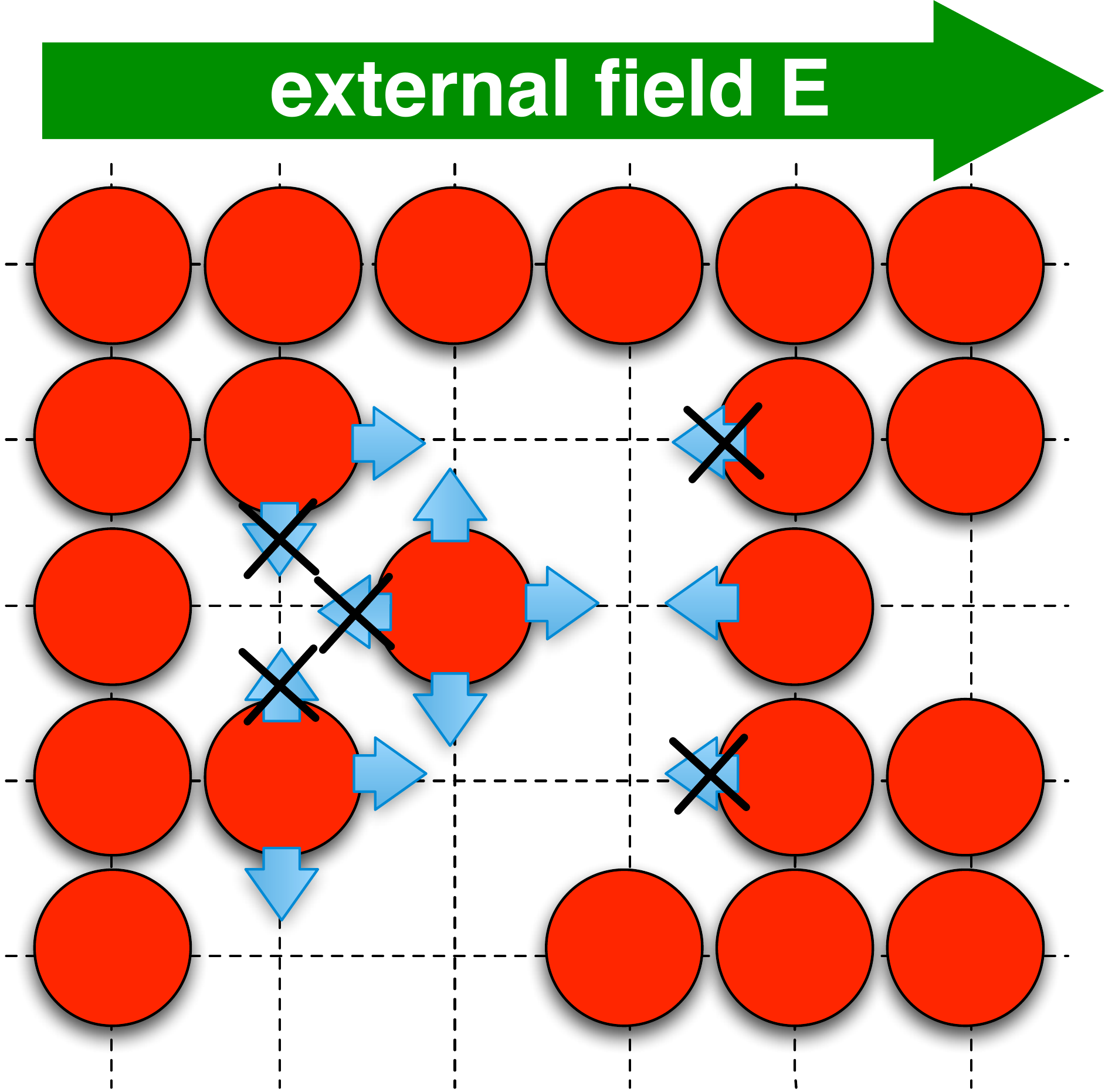}
\caption{(color online) Scheme of the model rules and the role of the kinetic
  constraints: particles (red) can move and occupy empty sites if they
  have at least 2 empty neighbors before and after the move. The
  external field is uniform and biases the movement of the particles
  from left to right reducing the probability of backward moves.}
\label{default}
\end{center}
\end{figure}

\section{The model}
\label{model}

We consider the model proposed by one of us in Ref.~\cite{sellitto}.
It can be viewed either as the kinetically constrained version of a
$2D$ Asymmetric Simple Exclusion Process (ASEP) or as the Kob-Andersen
model in presence of an external (non Hamiltonian) field.  In the
absence of drive the model has been largely studied; see,
e.g., Refs.~\cite{KoAn,KuPeSe,FrMuPa,ToBiFi,MaPi,ChSaKo,PaPiCaCo}.  We
take a $2D$ regular square lattice of size $L \times L$ with periodic
boundary conditions, in which the particle density $\rho$ is a
conserved parameter. The $N=\rho L^{2}$ particles are initially put at
random on the lattice, and an external field $E$ is applied along the
horizontal direction, from left to right, as illustrated in
Fig.~\ref{default}.  In this situation, the probability that a
particle moves against the field is $p_{back}={e}^{-E/\kB T}$,
while each other direction is equiprobable. We shall set the Boltzmann
constant $\kB$ to 1 throughout the paper and absorb the temperature
$T$ in the field definition, as the system we consider is purely
athermal and no additional energetic interaction among particles is
considered.

The model dynamics is fully described by the following steps:
\begin{enumerate}[(i)]
\item A particle is chosen at random uniformly.
\item The particle attempts to move along one of the four possible
  directions, by choosing one of its nearest neighbors site randomly
  with equal probability ($1/4$).
\item The particle motion to the randomly chosen site takes place only
  if the site is empty and the particle has at least 2 empty neighbors
  before and after the move. This latter condition is the so-called
  {\em kinetic constraint}.
\item If the previous condition is satisfied, the particle always moves provided
  that motion does not occur against the applied field; otherwise, if
  the particle attempts to move against the field, the motion occurs
  only if a random number uniformly chosen in the range $[0,1]$ is
  less than ${e}^{-E}$. This step is known as the {\em Metropolis
    rule}.
\end{enumerate}
We measure time in unit of Monte Carlo sweep, corresponding to the
random sequential update of the state of each particle on average.
Using a Metropolis-like algorithm allows one to make contact with some
standard results found in the literature on the ASEP and the Totally
Asymmetric Simple Exclusion process (TASEP), which are recovered here
in the absence of kinetic constraints and infinite field.  The
Metropolis choice maximizes the number of moves in the field
direction, because every attempt to move a particle along the field in
the unit time and by a lattice spacing is always accepted. It is
however not evident \textit{a priori} that the transport properties are
qualitatively independent from the chosen evolution rule.  Therefore
we have implemented, as an alternative to the Metropolis algorithm, a
Glauber-type dynamics according to which the probability to move a
particle against or along the applied field depends on whether the
random number, uniformly chosen in the range $[0,1]$, is less or
larger than $(1+{e}^{E})^{-1}$, respectively.  Results are pretty
robust and confirm our expectation that the transport properties we
found are generic: they are essentially due to the presence of kinetic
constraints that cannot be violated, no matter the choice of
transition probabilities. Finally, we notice that the local time
reversibility of the microscopic dynamics is satisfied.

\section{Transport regimes: current vs field relation}
\label{current}

The central quantity we focus on in this section is the particle
current, $J$, which is defined as the number of jumps in the field
direction minus the one in the opposite direction per lattice site and
per unit time.  It allows a first macroscopic characterization of the
different transport regimes present in the system. We generally
observe the existence of a threshold density $\rhoc \simeq 0.79$,
below which the current vs field relation is monotonic and above which
the current exhibits a crossover from a linear (ohmic) regime to a
negative differential resistance (non ohmic) regime at increasing field; see
figs. \ref{currdens} and \ref{saturating}.

\subsection{Low density regime}

\begin{figure}[b]
\begin{center}
\includegraphics[width=\columnwidth]{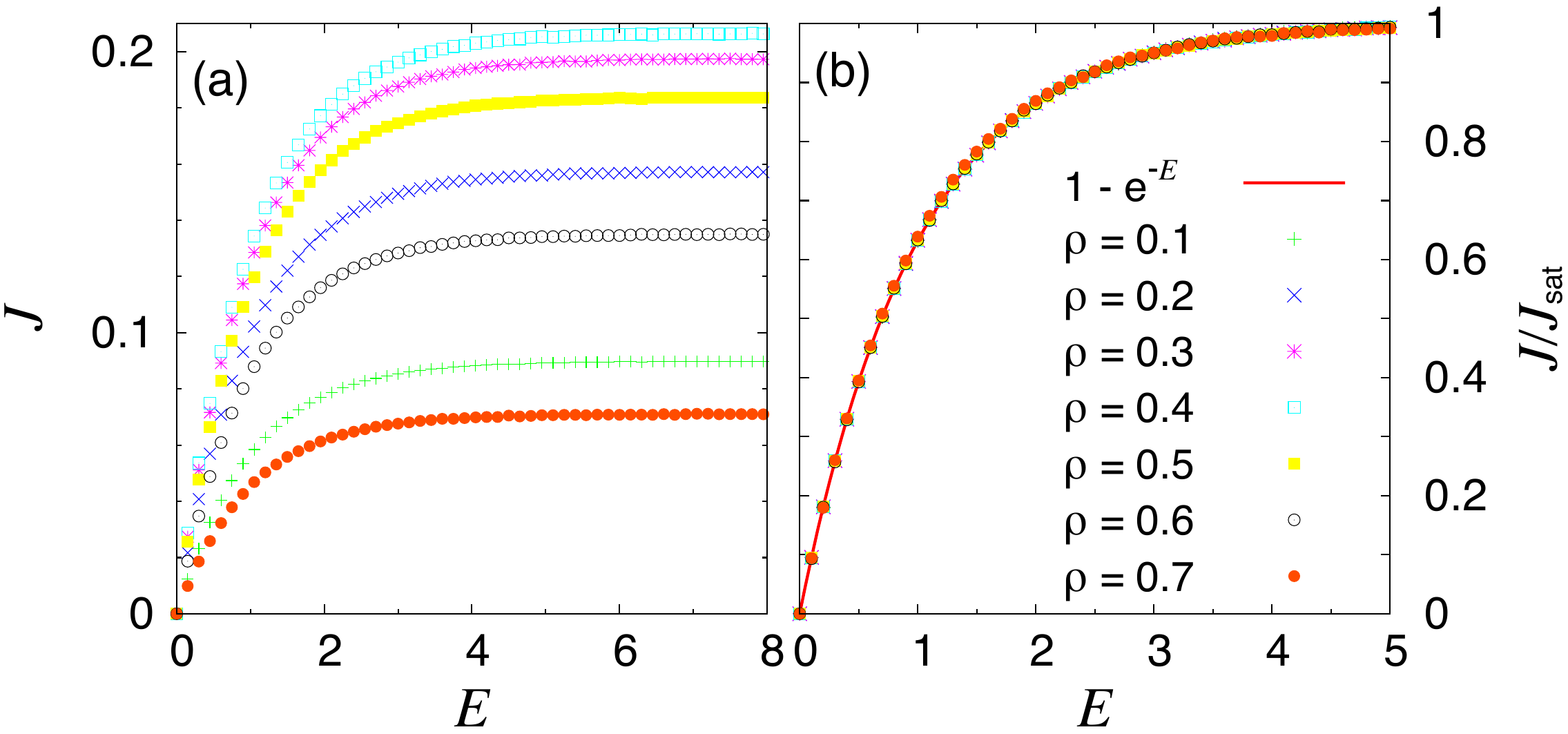}
\caption{(color online) (a) Current $J$ vs field $E$ relation for
  subcritical densities, $\rho < \rhoc \simeq 0.79$. The system
  size is $L^2=50^2$. The current is a monotonic function of the field,
  saturating at large field. Its behavior is qualitatively similar to
  the ASEP on a 2D square lattice.  (b) The density dependence of the current vs
  field relation can be easily accounted for by rescaling $J(E)$ with
  the saturation current $J_{\rm \scriptstyle sat}(\rho)$. The current
  ratio $J/J_{\rm \scriptstyle sat}$ is exactly equal to the difference
  between forward and backward transition probabilities, $1-{e}^{-E}$.}
\label{currdens}
\end{center}
\end{figure}

In the small density regime, we expect that transport is weakly
influenced by the presence of kinetic constraints: indeed numerical
simulations show that the current vs field relation has a form much
similar to the ASEP~\cite{asep}; see Fig.~\ref{currdens}. So we can
set:
\begin{equation}
  J(\rho, E)= A \, \frac{1}{4} \rho (1-\rho) (1-{e}^{-E}),
\label{JvsE-smallrho}
\end{equation}
with the pre-factor $A$ accounting for a further possible dependence
on $\rho$ and $E$ ascribed to the constrained dynamics.  We can
consider two limiting cases. In the absence of constraints, the
pre-factor $A$ must be 1, consistently with the ASEP.  When the
field becomes very large, the current saturates to a finite value,
which for the standard TASEP is $J_{\rm \scriptstyle sat} \sim \rho (1-\rho)$.  When
increasing the particles density the effect of the constraints is to
reduce the number of accessible paths in the configuration space and
to slow down the dynamics so that the current is smaller than what
expected in the unconstrained case. Interestingly enough, even though
the value of the pre-factor $A$ decreases continuously with increasing
$\rho$, it does not depend on the applied field, so that
Eq.~(\ref{JvsE-smallrho}) takes the same scaling form in the whole
$\rho < \rhoc \simeq 0.79$ regime, as far as its field dependence is
concerned; see Fig.~\ref{currdens}(b).

\begin{figure}[b]
\begin{center}
\includegraphics[width=\columnwidth]{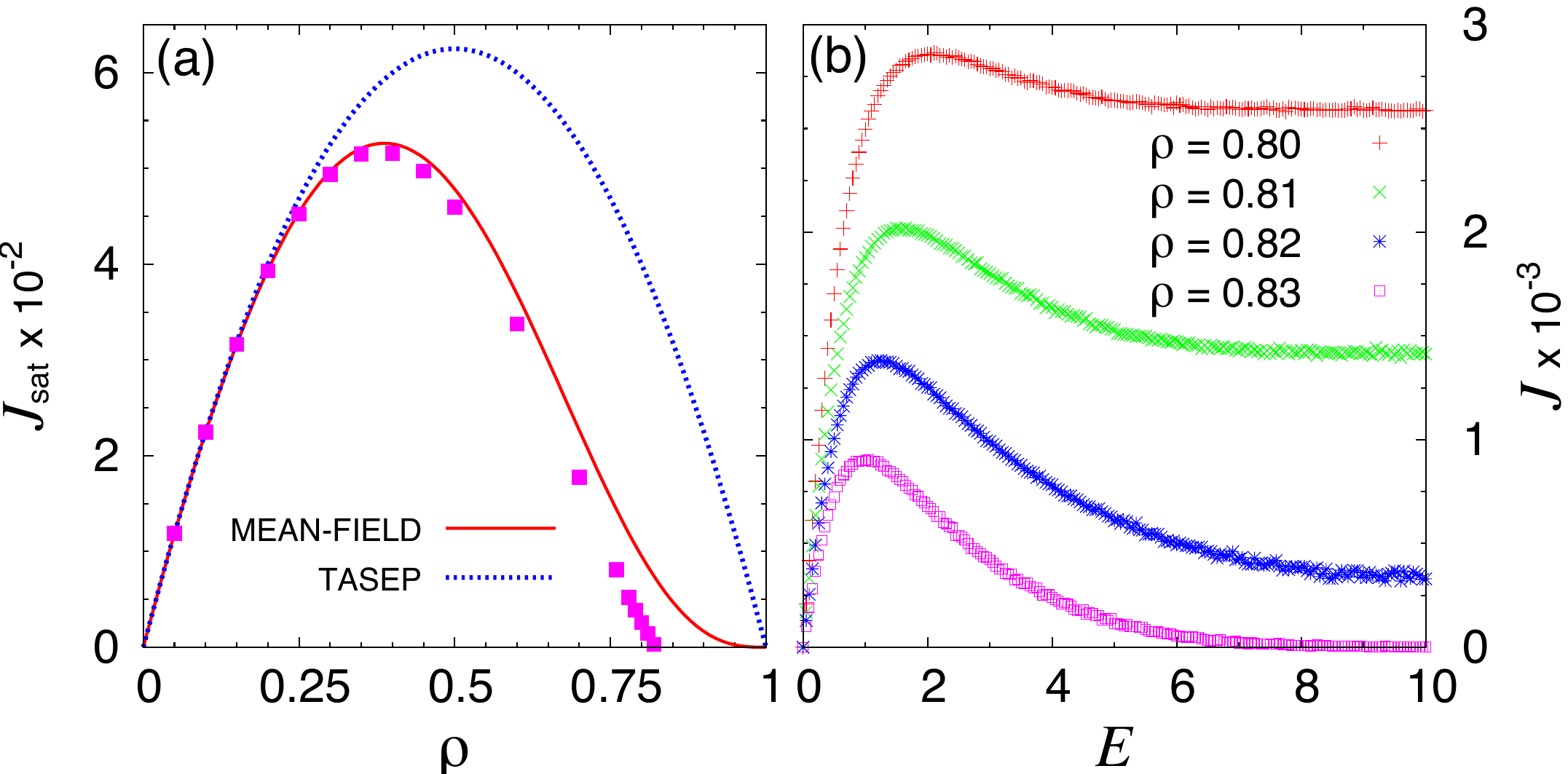}
\caption{(color online) (a) The current as a function of the density
  for our model with the kinetic constraint in the limit of very strong
  fields.  The peak is at $\rho\approx 0.4$ while for the classical
  TASEP (blue dashed line) the peak is at $\rho=0.5$. The mean-field
  formula (\ref{MF}) (red full line) shows a good qualitative agreement
  that becomes quantitative at densities below the peak value. (b) The
  most surprising property of the model: at high densities, $\rho >
  \rhoc \simeq 0.79$, the simulations provide a current-field
  relation which is non-monotonic, as discussed in
  Ref.~\cite{sellitto}.  Simulation data at $\rho > \rhoc$ are
  obtained by using systems of size $L^2=400^2$.}
\label{saturating}
\end{center}
\end{figure}

The similarity between the two behaviors - with and without
constraints - suggests that the density dependence of $A(\rho)$ can be
estimated by the means of a mean-field approach that neglects the role
of two-point and higher-order correlations.  This approximation allows
us to quantify the value of $A(\rho)$, writing
\begin{equation}
A(\rho)=(1-\rho^{3})^{2},
\end{equation}
which implies
\begin{equation}
  J(\rho, E)=\frac{1}{4} (1-{e}^{-E})\rho (1-\rho)
  (1-\rho^{3})^{2} .
\label{MF}
\end{equation}
The analysis of this expression is straightforward: the factor $1/4$
accounts for the four possible directions of motion on the $2D$ square
lattice; the term $1-{e}^{-E}$ is the difference between the
forward and backward transition probabilities; the product
$\rho(1-\rho)$ gives the probability to find a particle on a certain
site of the lattice with a nearby hole, if all correlations are
neglected; in this approximation, the last term $(1-\rho^{3})^2$
simply accounts for the kinetic constraint: It reads as the probability
to have at least two empty neighbors (which is equal to that one of
not having three occupied neighbors), and is counted twice because of
the local microscopic reversibility of the kinetic rule.
In the strong field limit, $E \to \infty$, the current saturates to
the value
\begin{equation}
  J_{\rm \scriptstyle sat}(\rho)=\frac{1}{4} \rho (1-\rho)
  (1-\rho^{3})^{2} 
\label{MFsat}
\end{equation}
As shown in Fig.~\ref{saturating}(a), the mean-field approximation for
the saturation current works well for small densities, suggesting the
higher order correlations are negligible in that regime. When the
density of particles increases, larger and larger correlations appear
and above a certain density the mean field approach breaks down.
Notice that in spite of the local microscopic time-reversibility of
the kinetic rule and particle-hole symmetry, the interplay of the
driving force and the kinetic constraints leads, in the limit of very
strong fields, to an asymmetric current vs density relation as shown
in Fig.~\ref{saturating}(a). The emergence of global particle-hole
broken symmetry can be understood as follows: the two limit situations
where only one particle is present on the lattice and where there is
only one hole do not lead to the same current: the first case has a
finite current $J=\frac{1}{4}(1-{ e}^{-E})/L^2$ while in the second case the
current is strictly zero, due to the caging rules.  One can compare the constrained model
result at strong field with the TASEP, where the current is given by a
parabola peaked in $\rho=0.5$.  On the contrary, the constrained model
is peaked around a smaller density ($\approx 0.4$) and shows an almost
zero current region at particle density near 1.

\subsection{High-density regime}

At density above $\rhoc$, see Fig.~\ref{saturating}.(b),  the non-monotonic behavior of $J(E)$ emerges
as the signature of extra mechanisms producing a more complex
transport dynamics.  Such non-monotonic behavior is more and more
pronounced with increasing density and is related to the growth of
several orders of magnitude of the relaxation times of the system
\cite{sellitto}. At such high densities one can distinguish between
two dynamical regimes: a positive resistance regime, where the current
grows linearly with the field, and a negative resistance one, where
the increase of the field corresponds to a decrease in the current, see Fig.~\ref{saturating}.(b).
The occurrence of non-monotonic transport can be qualitatively
understood as a consequence of the decreasing probability of backword
motion: at high density and increasing field, the particle
rearrangements needed to remove obstruction to the flow, require more
and more particle moves against, or normal to, the field direction,
and this leads to a flow reduction.  In particular, three distinct
behaviors can be considered, as discussed in Ref.~\cite{sellitto}: (I)
$J_{\rm \scriptstyle sat}$ is finite; (II) $J_{\rm \scriptstyle sat}$ is
vanishingly small, if not zero; and (III) $J(E)$ vanishes above a
finite driving force, $E > E_{\rm \scriptstyle c}$. Numerical results
suggest that regime I occurs in the range $\rhoc <\rho<0.83$ while
regime II appears at a higher density. The existence of the jamming
regime III cannot be obviously ascertained due to the strong
finite-size effects related to bootstrap percolation. The
characterization of these effects is notoriously difficult and so this
jamming regime will not be discussed here.  Rather, the main subject
of this work will be the crossover between the linear and the
non-monotonic transport regimes for a moderately large field and not too
high particle densities.

\section{Space and time heterogeneities}
\label{sptm}

\begin{figure}[t]
\begin{center}
\includegraphics[width=\columnwidth]{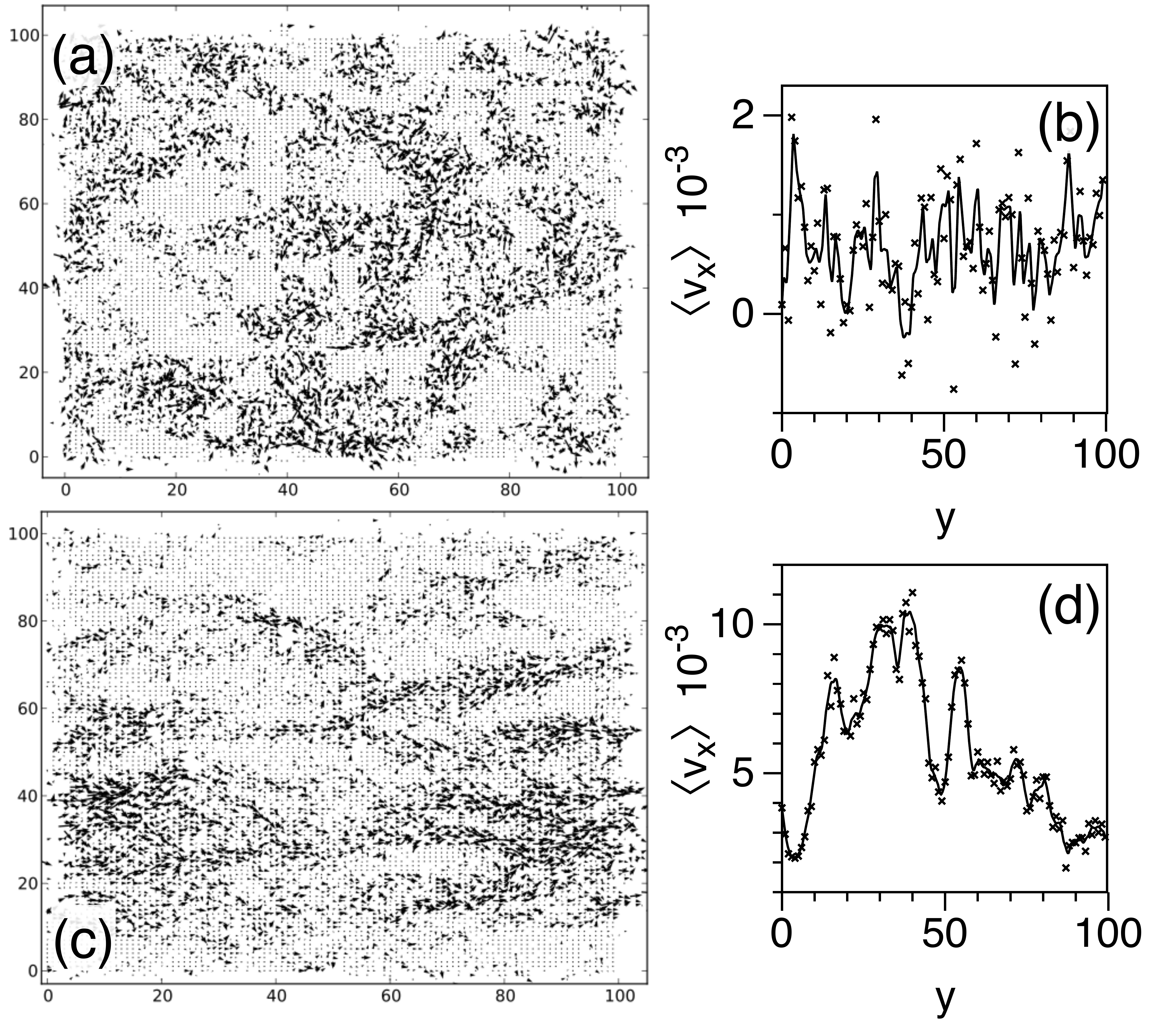}
\caption{Average velocity field for an $L=100$ system at density
  $\rho=0.80$, with $E$ oriented left-to-right. In the top panels (a)
  and (b) the system is in the linear regime, $E=0.1$, while in the
  bottom panels (c) and (d) the system is in the negative differential
  resistance regime, $E=2.8$.  The time window considered for the
  evaluation of the average speed is $t_{w}\approx 1 \taurel$.  At
  small fields the dynamics is homogeneous, while in the negative
  resistance regime, shear bands appear. The boxes (b) and (d) represent
  the longitudinal projections of the velocity vectors averaged over
  each horizontal line of particles as a function of the transversal
  coordinate $y$. }
\label{quivermap}
\end{center}
\end{figure}
\begin{figure}[t]
\begin{center}
\includegraphics[width=\columnwidth]{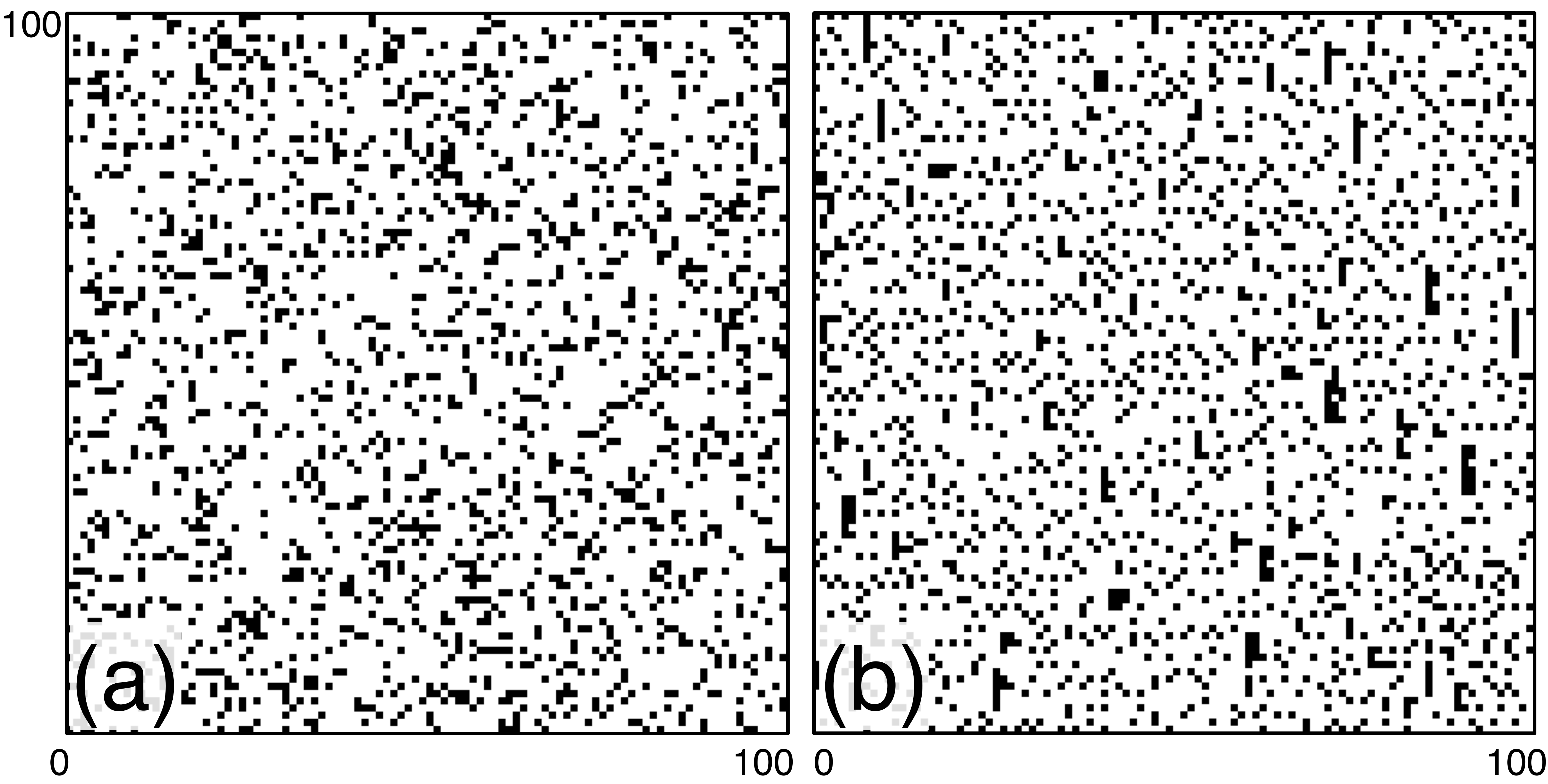}
\caption{Snapshots of two steady-state configurations with particles
  (white) and holes (black) for a system of size $L=100$ and $\rho=0.82$ at
  (a) $E=0$ and (b) $E=5$ with $E$ being in the horizontal direction
  (left to right). One can notice the vertical (transversal to the
  field) structures arising in the strong field case (b).}
\label{confs}
\end{center}
\end{figure}

\begin{figure}[htbp]
\begin{center}
\includegraphics[width=\columnwidth]{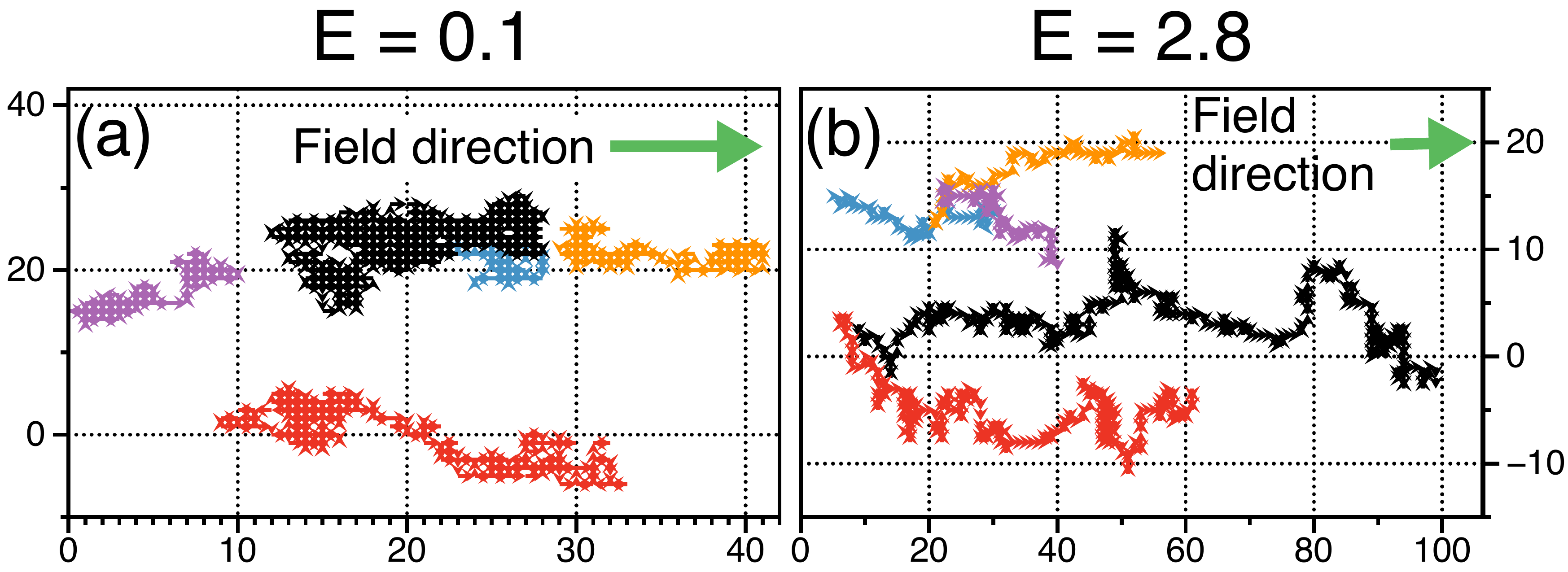}
\caption{(color oline) Some examples of real space particle
  trajectories of equal duration $\tau=10^{4}$ in the two current
  regimes for a system of density $\rho=0.80$: (a) at small fields, $E=0.1$, and (b) large fields, $E=2.8$, corresponding to a negative
  resistance behavior. One can observe the mainly diffusive, brownian
  behavior (a) vs the directed behavior (b) where many steps are spent
  in wandering moves in the transversal direction. Each arrow
  corresponds to a directed step and the axes are in lattice spacing
  units.}
\label{trajectories}
\end{center}
\end{figure}

The previous analysis suggests that the increase of the density beyond
the critical value $\rhoc$ corresponds to the switch between two
qualitatively different transport regimes. The first step for the
description of the high density regime is to analyze the configuration
space, looking for a direct relationship between the non-monotonic
behavior of $J(E)$ and changes in the typical arrangements of the
particles, similarly to what has been done for other one-dimensional (1D) or 2D models
(see, for example, \cite{JKGC}).

By the means of direct inspection, one can observe that the increase
of the field leads to a transversal symmetry breaking in configuration
space: not only do particles form longitudinal flowing bands along the
field direction (see figure \ref{quivermap}), but one can already
visualize emerging structures composed by blocked and empty regions,
inducing an intermittent dynamics for particles
(fig.~\ref{confs}). Particles are trapped for very long times,
wandering diffusively in the transversal direction, and occasionally
make a fully directed jump in the field direction
(fig.~\ref{trajectories}): such a behavior is responsible for the
anomalous diffusion observed in previous works \cite{sellitto}.

The inhomogeneities in the dynamics start to appear when the $J(E)$
peak is crossed and correspond to a coexistence between blocked and
mobile trajectories.  The average velocity field over time windows
below the relaxation time of the system allows for a proper
representation (fig.~\ref{quivermap}).  Longitudinal bands of
different mobility can be seen at large fields, while at small fields
the spatial distribution of the velocity vectors is homogenous.  In
analogy with what is observed for sheared systems
(i.e., \cite{shearbands}), we can call such structures \textit{shear
  bands}. Nonetheless, these bands are not localized within the
system, but have an intermittent and transient nature since the system
is homogeneous if averaged over sufficiently long times. During the
evolution, all the particles belong both to active and inactive bands.

\section{Anisotropic space correlations}
\label{corr}

Since the driven dynamics is obviously non-isotropic, we investigate
here several measures of spatial anisotropy, namely, transversal and
longitudinal persistence, dynamic susceptibility, two-point
correlations, and the van Hove intermediate self-scattering function.

\begin{figure}[t]
\begin{center}
\includegraphics[height=0.85\columnwidth]{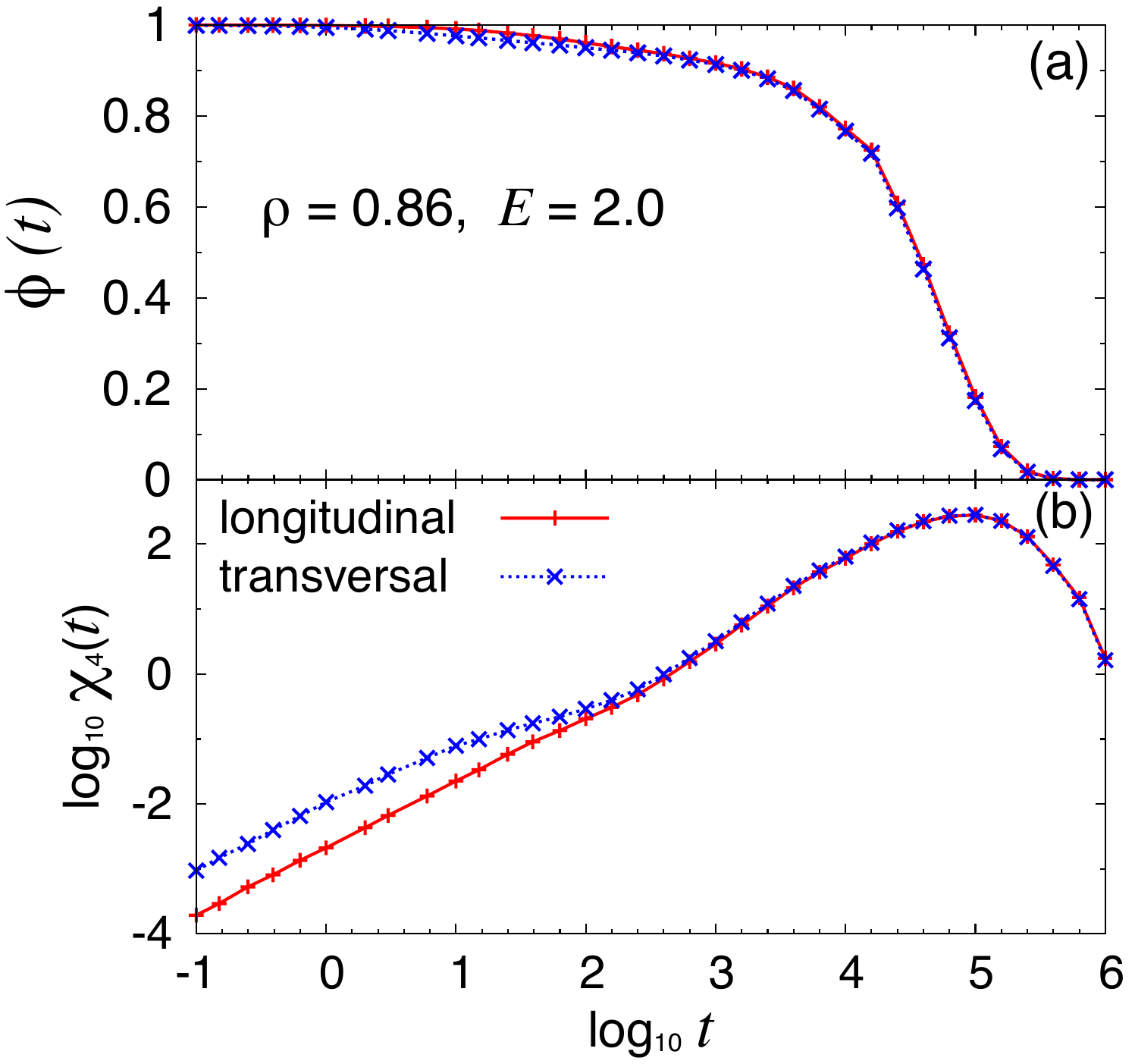}
\end{center}
\caption{(color online) (a) Transversal ($\times$) and longitudinal (+) persistence,
  $\phi(t)$, vs time, $t$, for particle density $\rho=0.86$ and
  applied field $E=2.0$, corresponding to the negative differential
  resistance regime. Square lattice of linear size $L=500$. (b)
  transversal and longitudinal persistence fluctuations, $\chi_4$, for
  $\rho=0.86$ and $E=2.0$. Square lattice of linear size $L=100$.}
\label{Fig:g_rho0.86pers}
\end{figure}

\subsection{Persistence and dynamic susceptibility}

It is customary to characterize the dynamics of kinetically
constrained systems by the persistence function, $\phi(t)$, i.e., the
probability that a particle has never moved between times $0$ and
$t$, whose field and density dependence have previously been discussed in Ref. \cite{sellitto}. The long-time limit of $\phi(t)$ represents the fraction of
particles that never moved, i.e., the fraction of permanently
blocked particles. An asymptotic finite value of $\phi(t)$, therefore,
signals a transition to a dynamically broken ergodicity regime.  In
our anisotropic system, we obviously need to distinguish between
transversal and longitudinal particle motion leading to the definition
of $\phiT(t)$ and $\phiL(t)$.

We find that the difference between longitudinal and transversal
persistence functions is not sizable at a small field, and tiny at larger
fields. In particular, it is only apparent in the early stage of
relaxation of the large field regime.  
This suggests that there are no long-lived correlated structures but
rather the continuous creation and destruction of spatially extended
defects facilitating particle transport. Clearly, since persistence
is a global, time-integrated quantity, it cannot represent an accurate
probe of dynamic anisotropy on short-time scales. A slightly better
characterization is provided by the dynamic susceptibility, which is
generally defined as mean-square fluctuations of persistence
\begin{equation}
\chi_4(t) = N \left( \langle \phi^2(t)\rangle - \langle \phi(t)
\rangle^2 \right).
\end{equation}
In Fig. \ref{Fig:g_rho0.86pers} we plot the transversal and longitudinal
component of persistence fluctuations. Differences between the two
components are now more clearly visible at early times, and suggest
the formation of short-lived correlated structures in the transversal
direction. On a longer timescale the two susceptibility components
show similar behavior: both the peaks position and the peaks height
coincide, so that the behavior of the relaxation time as measured from
the susceptibility peaks remains essentially unchanged.

\begin{figure}[t]
\begin{center}
\includegraphics[height=0.85\columnwidth]{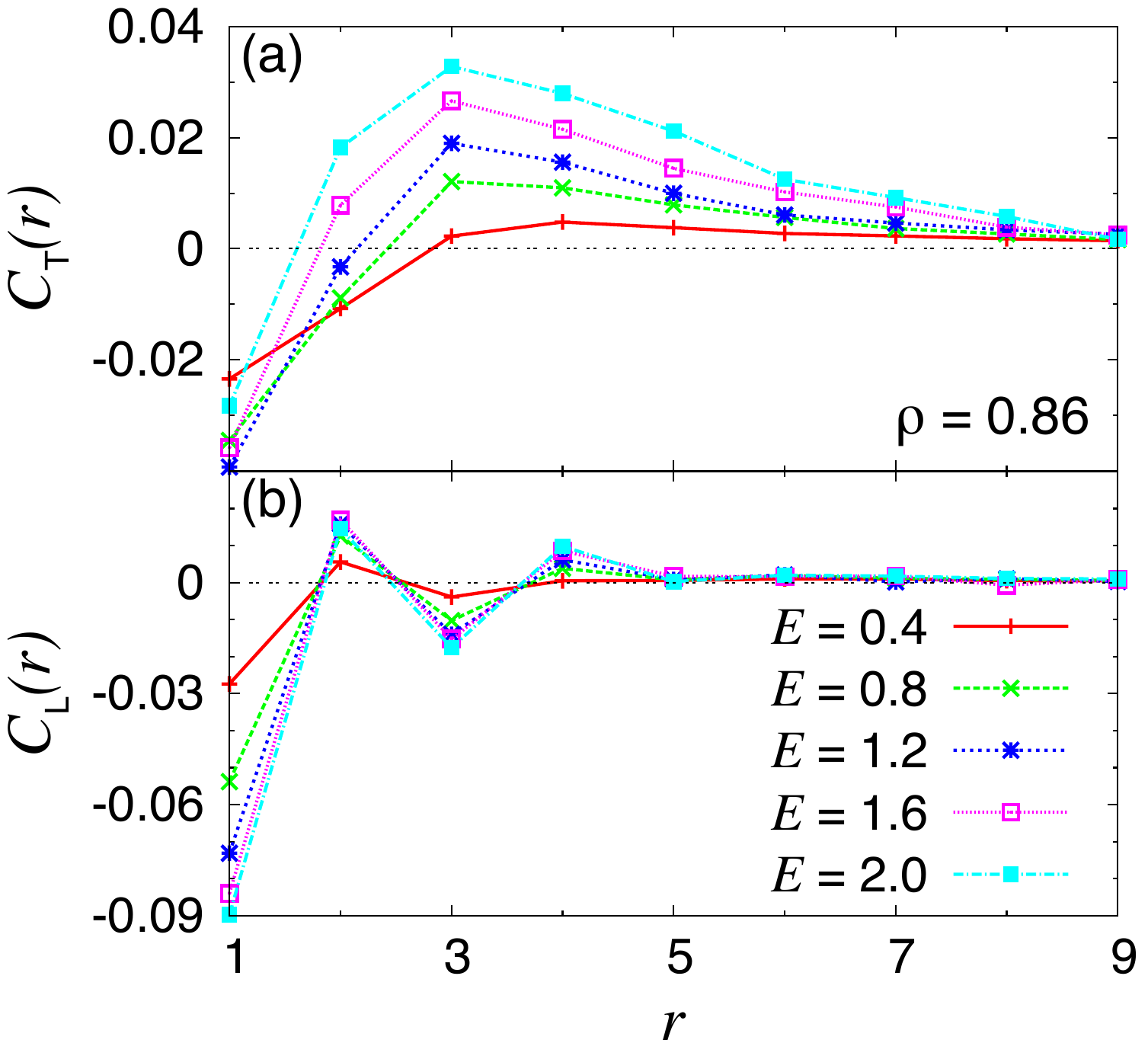}
\
\end{center}
\caption{(color online) (a) Transversal  and (b) longitudinal two-point
  correlation functions for $\rho=0.86$ for several applied fields $E$.
  Square lattice of linear size $L=500$.  }
\label{Fig:g_rho0.86}
\end{figure}

\subsection{Two-point correlation}

To better quantify the anisotropy of dynamics we investigate here the
behavior of a two-point correlation function at various values of
density and driving field.  The two-point correlation $C({\bf r})$
function is a measure of the spatial correlation of two particles at
distance ${\bf r}$, and is defined here as
\begin{equation}
C({\bf r}) = \frac{\left[ \langle n_{\bf r+r_0} n_{\bf r_0} \rangle
    \right ]_{\bf r_0} - \rho^2}{\rho(1-\rho)} ,
\end{equation}
where the square brackets denote a spatial average and $n_{\bf r}=0,
1$ are the usual lattice-gas occupation variables.  In
fig.~\ref{Fig:g_rho0.86} we show the transversal, $\gT({\bf r})$, and
the longitudinal, $\gL({\bf r})$, two-point correlation functions in
the non-equilibrium steady state.

\begin{figure}[h]
\begin{center}
  \includegraphics[width=\columnwidth]{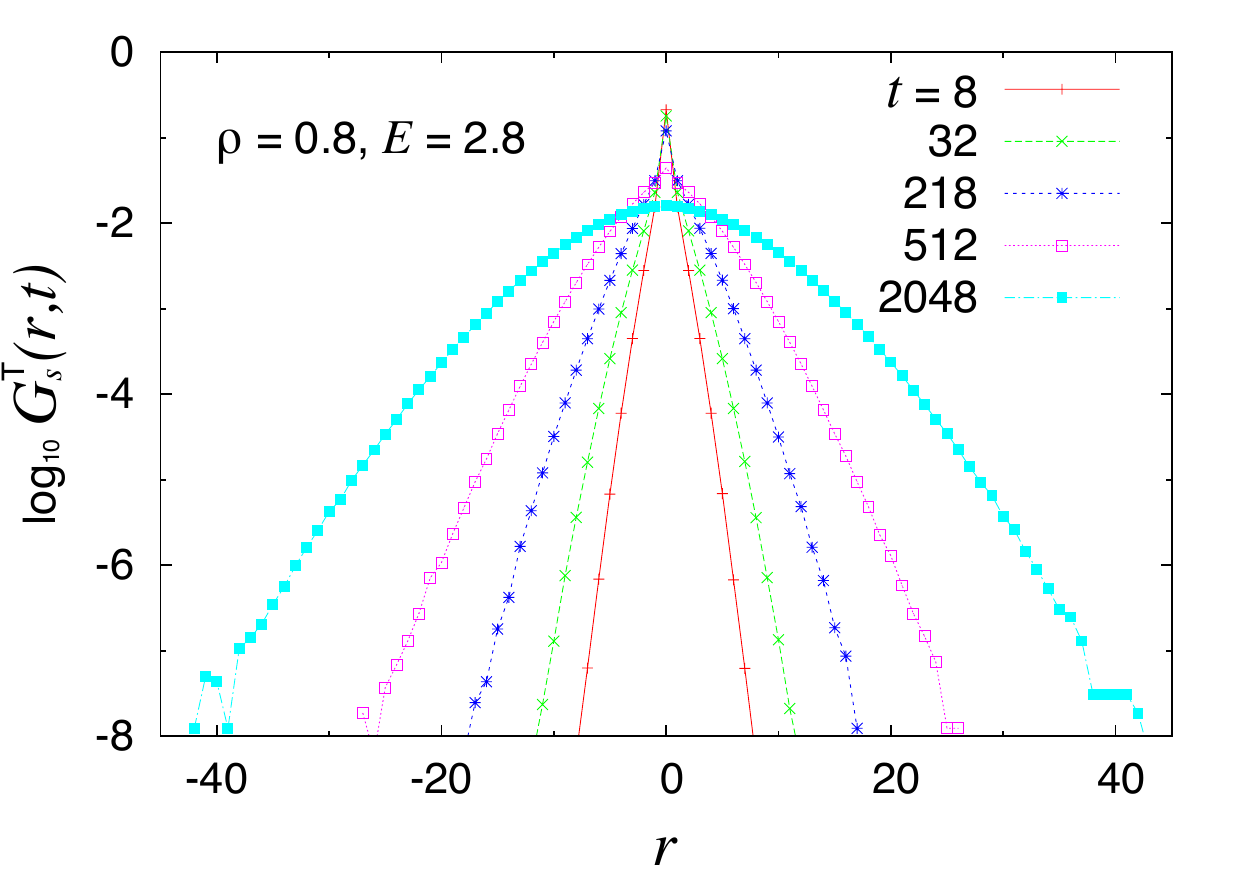}
\end{center}
\caption{(color online) Transversal part of van Hove self correlation function at
  particle density $\rho=0.8$ and field $E=2.8$ (negative resistance
  regime). The system consists of a square lattice of linear size
  $L=50$.}
\label{Fig:self2}
\end{figure}

In spite of the effective hard-core-like repulsion generated by
kinetic constraints, we see that there is actually a medium short-range
attractive-like interaction in the transversal direction to the applied
force [fig.~\ref{Fig:g_rho0.86}(a)].  The increased correlation between
two nearby particles at a distance $r$, especially in the transversal
direction, can be qualitatively explained as a purely dynamic effect,
which arises from the fact that at large density and large applied
field, any particle needs first to move either backward or
transversally to the field direction in order to proceed forward.
This effect is more and more pronounced as the density and the applied
field increase.  The interplay of kinetic constraints and driving
force thus generally enhances the clustering of particles and appears
to be akin to a transversal static short-range attraction.  In the
longitudinal direction instead we observe a short-range oscillatory
behavior typical of liquids [fig.~\ref{Fig:g_rho0.86}(b)].  These
features can be linked to the different spatial structures that
actually exist in the transversal and the longitudinal direction: we
describe and discuss this in section~\ref{traps}.

\begin{figure*}[!ht]
\begin{center}
\includegraphics[width=\textwidth]{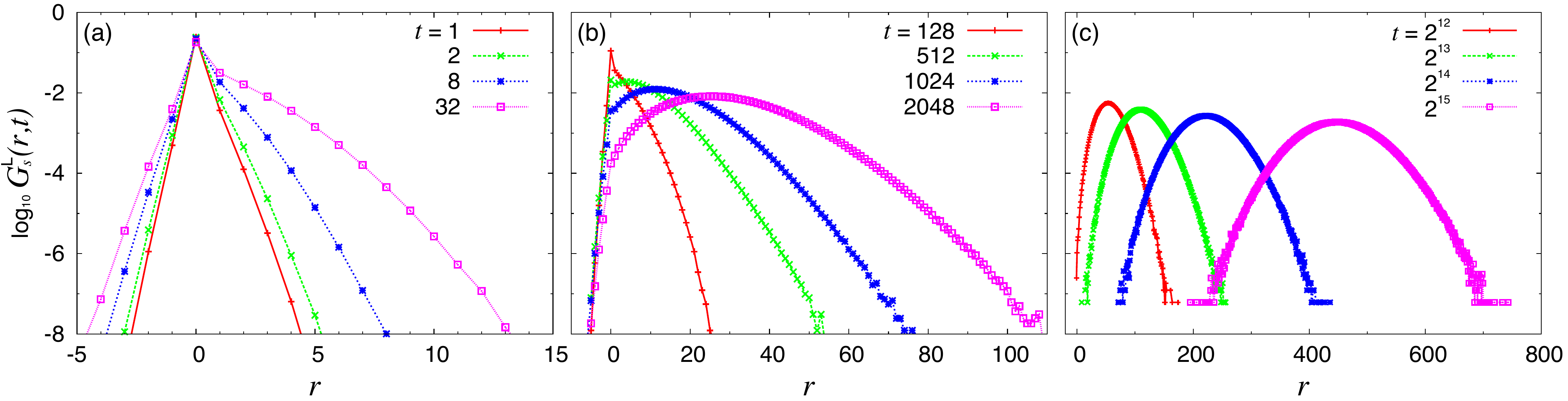}
\end{center}
\caption{(color online) Longitudinal part of van Hove self-correlation function
  $G_s(r,t)$ vs position $r$ at time $t$, particle density $\rho =0.8$,
  and field $E=2.8$ (negative resistance regime) for a system of
  linear size $L=50$. There is long-lived asymmetry induced by the
  interplay of external drift and kinetic constraints. The exponential
  tails observed at short times eventually become more and more
  Gaussian at longer times.
  \hspace{-0.4cm}
}
\label{Fig:self1}
\end{figure*}

\subsection{van Hove self-correlation function}

The van Hove self correlation function $G_s(r,t)$, quantifying  the probability that a particle makes a displacement of size $r$ over a time interval $t$, is defined as
\begin{equation}
  G_s(r,t) = \frac{1}{N} \sum_{i=1}^N \left \langle 
  \delta ( |{\bf r}_i(t) - {\bf r}_i(0)| - r)  \right \rangle
\end{equation}
 where the delta is the Kronecker delta function.  
When the motion of particles is diffusive 
the van Hove function takes a Gaussian form. 
Deviations from the Gaussian behavior have been observed in a variety
of glassy systems. Typically, one finds a crossover from an
exponential decay at short-time, which is suggestive of dynamic
heterogeneities (some particles move faster than others) to a
Gaussian, normal diffusive behavior at large time - see Figs.
\ref{Fig:self1} and \ref{Fig:self2}.  Consistently with other studies
of relaxational glassy dynamics \cite{Chaudhuri,Stariolo}, we find a
similar behavior in the particle transversal motion of our system.  The
latter, indeed can be assimilated to an equilibrium subsystem as there
is no violation of detailed balance in the transversal
direction. Interestingly, in the longitudinal direction, instead, we
observe an asymmetric distribution of particle motion at early times,
with Gaussian behavior slowly recovered at late times. The origin of
the asymmetry in the distribution is due to the interplay between the
drift caused by the field and the presence of kinetic constraints
hindering crowded motion especially in the backward direction (against
the field). The tail on the left side stays exponential over a longer
time than on the right side, because the backward events leading to
larger structural rearrangements are more rare. At large enough field
backward motion is so obstructed that the time it takes to approach
the Gaussian behavior can be exceedingly long to be observed.


\begin{figure*}[!ht]
\begin{center}
\includegraphics[width=\textwidth]{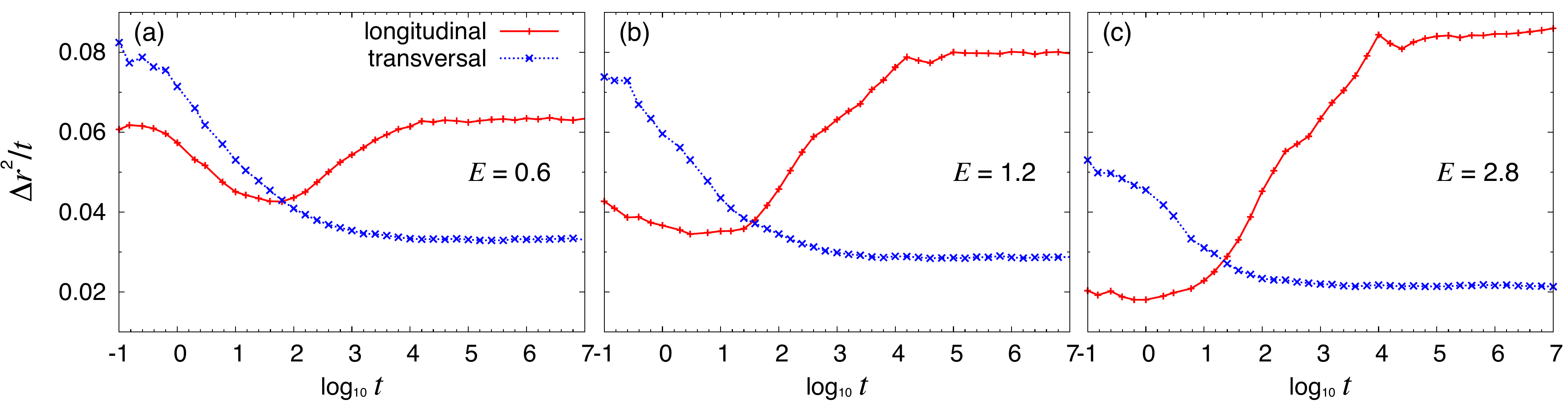}
\end{center}
\caption{(color online) Time averaged longitudinal (+) and transversal ($\times$) mean-square
  displacement $\Delta r^2/t$ for a square lattice of linear size
  $L=50$. The normal diffusive behavior $\Delta r^{2}\sim t$
  corresponds to horizontal lines; negative (positive) slope corresponds
  to sub (super) diffusion regime, respectively.
}
\label{Fig:msd}
\end{figure*}

\subsection{Mean-square displacement: anomalous diffusion}


The differences we observed in the longitudinal and transversal motion
are further confirmed by the analysis of the mean-square displacement.
In fig.~\ref{Fig:msd} we show the transversal and longitudinal
mean-square displacements as a function of time.  In the early stage
of the dynamics, we see a sub-diffusive behavior in the transversal
direction which correspond to the slow structural rearrangements of
small size. This regime shrinks at large field and the asymptotic
normal diffusion behavior is characterized by a diffusion coefficient
that decreases with the applied field.  In the longitudinal direction,
the initial short-time sub-diffusion is followed by an intermediate
super-diffusive behavior whose lifetime increases with the applied
field. It corresponds to the regime in which the longitudinal van Hove
function is strongly asymmetric and there are longitudinal particle
rearrangements of large size. Normal diffusion is recovered at late
times and, perhaps surprisingly, it is enhanced by increasing the
applied field.  We will show later that anisotropies are crucial for
the dynamics and can be described microscopically in terms of
intermittent creation and destruction of domain walls.

\section{Traps and domain walls}
\label{traps}

Although the space and time averaged macroscopic observables exhibit
several interesting features observed in more realistic systems, they
shed little light on the microscopic mechanisms responsible for the
blocking phase.  Several transport problems showing reduced mobility
involve the presence of localization and trapping of the carriers
\cite{JKGC,shlovskii,bouchaud-anom-diff,bouchaud-traps,Dhar-Dietrich}. In
these problems anomalous diffusion and broad distributions of the
waiting times of the particles are often found along with a
non-monotonous dependence of the particle current on the external
forcing or bias.  In this context, we want to relate $J(E)$ to some
specific properties of the domain walls or ``walls of holes'' that act
as trapping and blocking regions for the dynamics.

\begin{figure}[h!]
\begin{center}
\includegraphics[width=0.9\columnwidth]{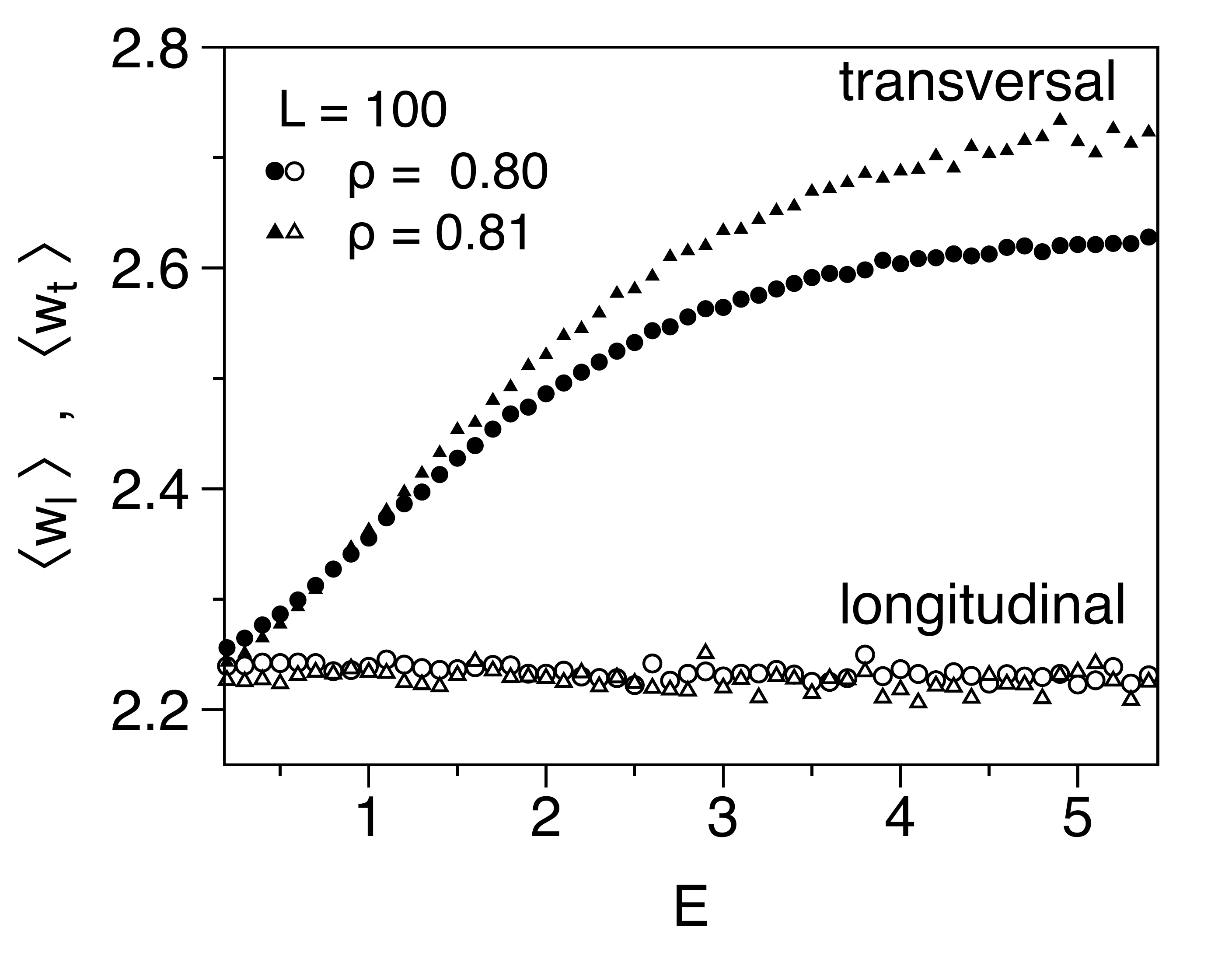}
\caption{Longitudinal $w_{l}$ (open symbols) and transversal $w_{t}$
  (filled symbols) average wall sizes for different values of the
  density as a function of the external forcing. The effect of the
  field on the longitudinal size is negligible with respect to the
  effect on the transversal size. Moreover, the higher the density,
  the stronger is the effect. }
\label{sizes}
\end{center}
\end{figure}

We have quantified these regions by the average value of their
longitudinal $w_{l}$ and transversal $w_{t}$ sizes. To do so, we have
chosen to define the walls in the simplest way: we decouple the
computation in the two directions and count any contiguous region of
at least 2 holes as a wall in the given direction. We find that the
direction sensitive to field intensity variations is the transversal
one (see fig. \ref{sizes}) reflecting the formation of the extended
structures that the direct inspection of the configurations already
suggested.  Nothing similar exist in the longitudinal direction.  For
this reason, we have concentrated our study on the transversal
direction, and considered the transversal size of the domain walls as
the key quantity to explain the negative resistance regime of $J(E)$;
we will name it $w:=w_{t}$ for simplicity. The interesting feature of
the growth of the average transversal size of the walls is that it is
a saturating function of the external field and allows for an
interpretation in terms of a blocking probability which will be detailed
in the following discussion.




\subsection{Origin of the walls}
\begin{figure}[t]
\begin{center}
\includegraphics[width=0.87\columnwidth]{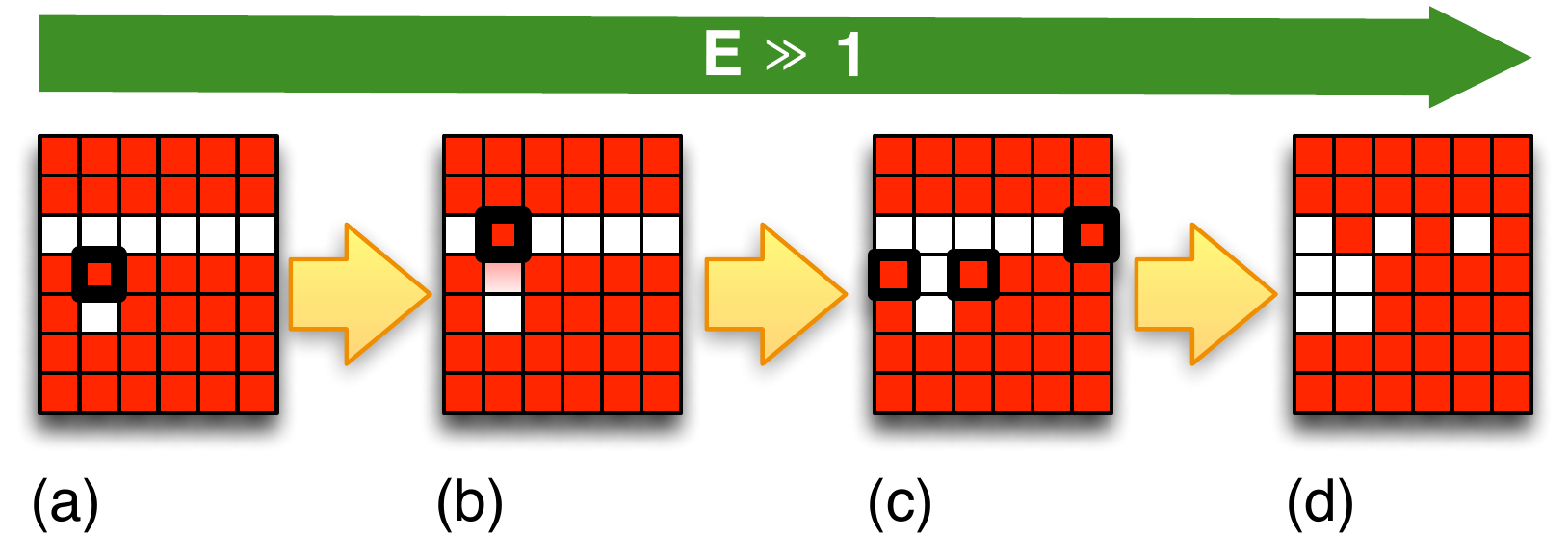}
\caption{(color online) At high densities and strong fields, longitudinal clusters of holes (empty squares) are easily reduced to transversal structures and basins. (a) The limit case of a single longitudinal strip of vacancies accessed by a mobile particle (bold bordered) is shown; (b) the particle enters the strip, freeing some new mobile particles; (c) the initial particle is pushed rightward by the field; and finally (d) the remaining mobile particles follow an analogous path and a transversal basin of holes is formed.}
\label{origin}
\end{center}
\end{figure}

The emergence of domain walls can be explained by a brief analysis of
the detailed microscopic moves for a specific configuration, and then
extending the results to the general case. Let us consider our system
when subject to very strong fields $E\gg1$. In that case, the
probability to move against the field is almost suppressed while the
transversal direction lets the system be mixed with a diffusive
mechanism. Let us consider a special configuration formed by a density
region where a longitudinal domain of empty regions has a single
mobile particle on its borders, as shown in figure \ref{origin}.(a).
If we follow the possible movements of the mobile particle and we
consider that it cannot move against the field, we see that after a
time which depends on the diffusive vertical process the particle is
pushed rightward in the sense of the field.  Other particles become
mobile and they follow an analogous path, so that eventually the holes
are concentrated in a basin that has lost the original longitudinal
form in favor to a more transversal structure, similar to what was
directly inspected in the snapshots of the evolution.

From this brief discussion we see that a general mechanism for
the formation of the walls exists and depends actually on the probability
of reversal moves $p_{back}$.  We also recognize that, at high
densities, if we have wide and compact empty regions and adjacent wide
and compact filled regions, small ``impurities'' formed by isolated
empty sites play an important role in making specific particles mobile.

\subsection{Exponential distributions}
In order to properly justify the choice of the observable associated
with the formation of domain walls, we have computed the distribution of
the transversal wall sizes: the data have been collected letting a
system in its steady state evolve for ~20 relaxation times and
scanning the whole lattice at each Montecarlo step. Moreover, we
repeated the process for 150 samples, in order to smooth the
distribution. Clearly, any distribution of lengths extracted from a
simulation is affected by finite size effects (i.e. the tails of the
distributions are bounded by the system size) so we are interested in
large systems. We have chosen, for the majority of the results shown
in the following figures, $L=100$, given that the crossover length
obtained in \cite{ToBiFi} for the undriven Kob-Andersen
model ranges from $\Xi_{0.8}\approx 16$ to $\Xi_{0.82}\approx 21$ for
values of $\rho$ between $0.80$ and $0.82$.

\begin{figure}[t]
\begin{center}
\includegraphics[width=\columnwidth]{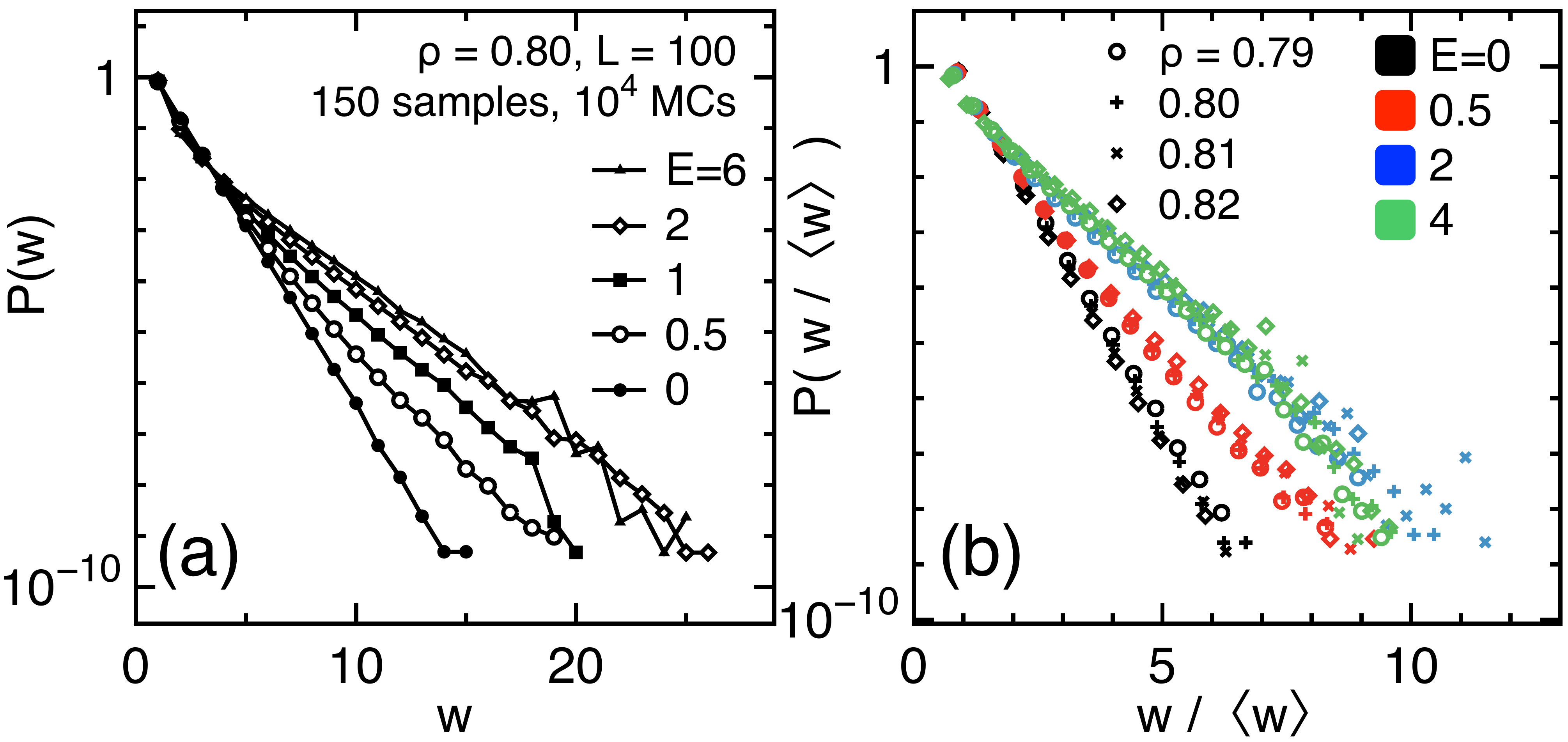}
\caption{(color online) (a) Distribution of the transversal wall sizes. Excluding
  the very small structures necessary to the diffusive dynamics, clear
  exponential tails are established, defining a typical transversal
  wall length. (b) Distributions of the transversal wall sizes for
  different values of the external field, rescaled with respect to the
  average walls size value at densities $0.79,0.80,0.81,0.82$. For
  each value of the external field, the distributions computed at
  different densities collapse and at a large field converge to the same
  distribution.}
\label{distr080}
\end{center}
\end{figure}

What we get in terms of the distributions is shown in figure
\ref{distr080} for the density $\rho=0.80$.  We plot also the
occurrence of very small structures of size 1 in order to show that
they are at odds with respect to the rest of the data points. Indeed,
the picture we get is that the distribution of transversal sizes has
an exponential form $P(w)\propto e^{{-w/\langle w \rangle}}$ in a wide region
bounded by the very small structures of size 1 and 2 (which are in the
limit of the definition of a wall itself) and the tail of very large
walls (which are rare and whose observation also depends on the system
size).  We see that at $E=0 $ such a distribution is very clearly an
exponential and also interpolates walls of size 1. 
As the field intensity increases, two behaviors emerge that correspond
roughly to the two current regimes: in both regimes, we have a large
number of very small structures, but the occurrence of larger walls
increases until a final distribution is reached which is almost the
same both for $E=2$ (around the current peak) and $E=6$ (far deep in
the negative resistance region).  Very similar distributions are
obtained for different densities, even if the $J(E)$ curves at
different $\rho$ have been shown to be quite different (see inset of fig.~\ref{currdens} for comparison): for this reason we have collapsed
all the distributions with respect to their average wall size, showing
that a common behavior exists for any density (see right panel of
fig.~\ref{distr080}).  The interesting feature of the exponential
distribution is that it properly defines a very pertinent observable,
the characteristic wall size $\langle w \rangle$ corresponding to the
average.  The value of the average depends on the region of
integration of the distribution: therefore, there is an important dependence
of the obtained value on the inferior limit of integration, given that
the upper bound corresponds to the size of the simulated system.  In our
case, we have chosen to ignore only the 1-site-long walls and compute
the averages as if the walls were well defined for any size $\geq 2$,
obtaining the result in figure \ref{sizes}.
  
\subsection{Current and traps}
\begin{figure}[htbp]
\begin{center}
\includegraphics[width=0.87\columnwidth]{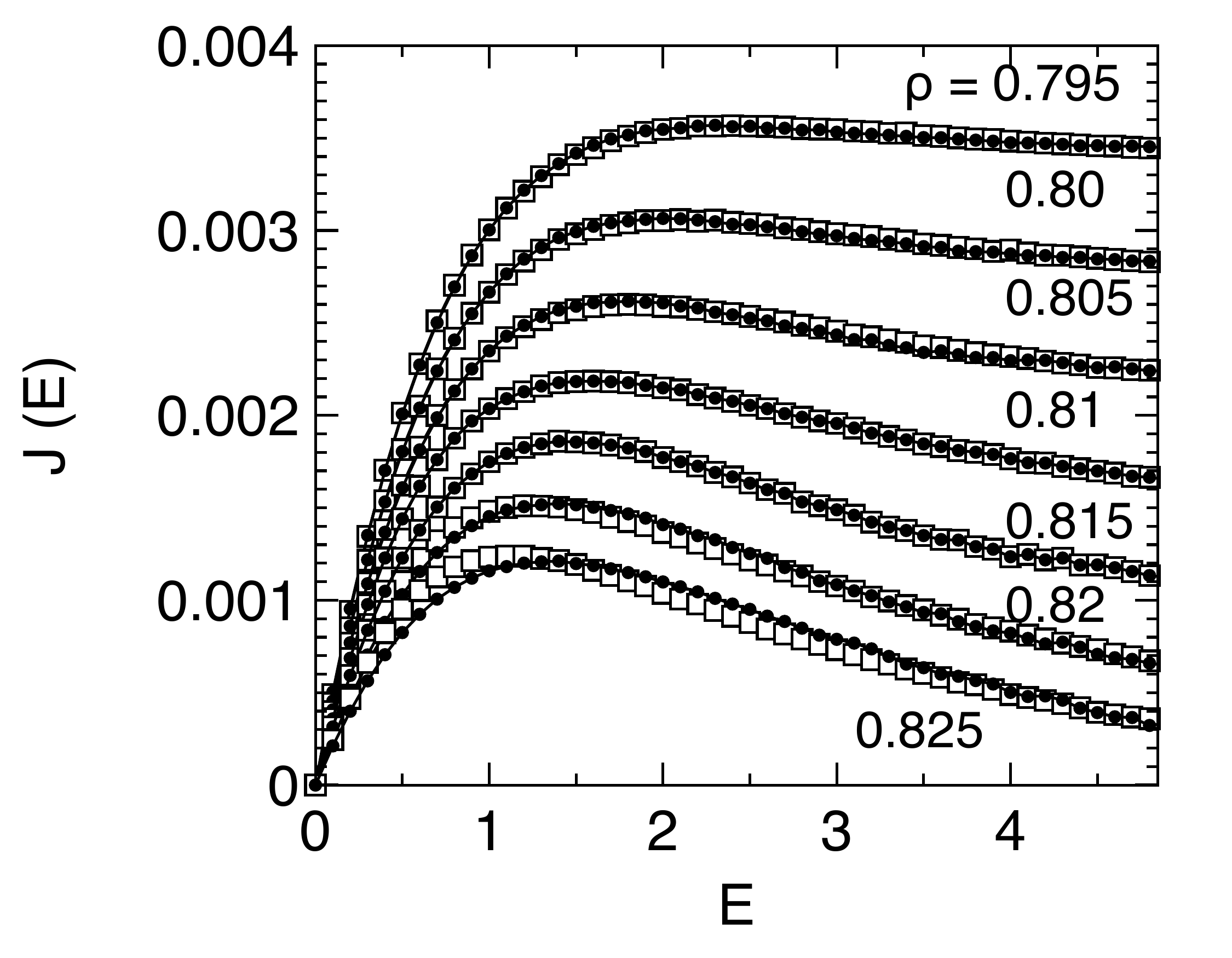}
\caption{Simulated flow curves (white squares) fitted by the
  phenomenological model $J(E) = A(1-e^{-E})(1-\alpha\langle w
  \rangle)$ (black dots). Densities range from $0.795$ to $0.825$, $L
  = 100$, and $\langle w \rangle$ is determined independently from the
  simulation.}
\label{fits}
\end{center}
\end{figure}

A phenomenological argument that allows for a fit expression of $J(E)$
with respect to $w(E)$ is the following. Let us say that the current
flowing into the system has the form
\begin{equation}
  J(E,\rho)=A(\rho)(1-{\rm e}^{-E})(1-p_{blocked}(E,\rho))
\end{equation}
where $p_{blocked}(E,\rho)$ is simply the probability to pick a
blocked configuration. We state that such a probability is expected to
be, at a first order of approximation, proportional to the average
transversal length of the domain walls.  We show this with a concise
reasoning: with a coarse grained view of the system, we can say that
each domain wall blocks a number of particle which is proportional to
its length.  If we suppose that the spacing between the different
walls $l_{w}$ depends only on the density, and we call $N_{w}$ the
total number of walls, we can say that the average number of blocked
sites is $n_{blocked}\approx N_{w} \langle w \rangle l_{w}$ and the
probability
\begin{equation}
  p_{blocked}\approx \frac{ N_{w} \langle w \rangle l_{w} }{\rho L^{2}}
\end{equation}
where $\rho L^{2}$ is the total number of particles.  Assuming that
the walls are uniformly distributed, we can estimate $N_{w}\propto
\frac{L^{2}}{l^{2}_{w}}$ and finally write that
\begin{equation}
  p_{blocked}\approx \alpha (\rho) \langle w \rangle(E,\rho) \propto
  \frac{\langle w \rangle(E,\rho)}{\rho l_{w}}
\end{equation}
where $\alpha(\rho)$ is only a fitting constant.  We obtain then an
empirical fitting expression
\begin{equation}
  J(E,\rho)=A(\rho)(1-e^{-E})[1-\alpha (\rho) \langle w \rangle(E,\rho) ]
\end{equation}
where both $A$ and $\alpha(\rho)$ should depend on the density of the
system and are pure fitting parameters.  This naive approach gives
good results for densities near the critical value $\rho_{c}=0.79$,
as shown in figure \ref{fits}, but fails for larger densities. It is
also harder to reach the steady state at high densities and fields and
properly compute the probability distribution of the wall sizes.
Nevertheless, this approach can explain on a phenomenological basis
the crossover region from the flowing to the blocking regime, with a
formal expression analogous to what was proposed in \cite{JKGC}, given
that $\langle w \rangle$ is a bounded growing function on the external
forcing.

\section{Conclusions}

We have investigated the spatio-temporal features of a simple
kinetically constrained model driven into a non-equilibrium stationary
state by a constant and uniform drive.  The model can be considered as
a generalization of the ASEP with an extra ingredient (the kinetic
constraints) schematically representing the cage effect of glassy
dynamics.  In this type of systems the interplay between kinetic
constraints and driving force generates some counterintuitive features
which are observed in more complex driven athermal systems, such as
highly packed colloidal suspensions and granular materials under
shear.  Despite the minimalistic rules of our model, we found a rich
transport behavior including a crossover from a linear-response
regime, at weak field and low density, to a negative resistance regime
at strong field and high density, and asymptotically broken
ergodicity.  We have shown that the flow reduction in the negative
resistance regime is related to the emergence of a complex
self-organization of dynamical structures which evolve intermittently
and exhibit anomalous diffusion.  Intermittency is due to the
competition between active and inactive regions, so that the system
evolves through the alternative succession of low and high mobility
configurations, compatibly with the scenario of a dynamical phase
transition suggested by the thermodynamics of histories
analysis~\cite{TurciPitard}.  As observed in other facilitated or
kinetically constrained models \cite{Jung:2004vo, JKGC,Shokef,Redner}, the
appearance of dynamical heterogeneities is accompanied by enhanced
diffusivity and is closely related to the intermittency in the
formation and disruption of particle clusters.  A systematic analysis of
the typical space and time averaged observables of complex liquids
shows that there are several interesting features associated with
transport. These include a dynamically induced short-range particle
attraction in the transversal direction, which is a signature of an
enhanced particle clustering, and a regime of super-diffusion behavior
in the longitudinal direction, whose duration increases at a larger
applied field. We have highlighted the intermittent and heterogeneous
nature of the dynamics by a careful analysis of the trajectories of
motion of the particles that actually contribute to the global
relaxation, and provided a phenomenological explanation of the
crossover to the negative resistance regime.  We have determined the
characteristic dynamical length of the system and its dependence on
the drift, connecting the detailed microscopic configuration space
structure to the macroscopic flow.  Further investigations concerning
the spatial distribution of the correlated walls of holes would
improve the understanding of the relationship between the two-folded
behavior of the current and the growth of $\langle w \rangle (E)$ with
increasing the field strength.

Several future developments can be envisaged. Extensions and
generalizations of the present model shall explore other
spatio-temporal features of non-equilibrium steady states. First, since
the dynamics at strong fields and high densities partitions the
systems in mobile and immobile dynamical regions, it would be
interesting to analyze how the mobility percolates through the system
and what are the geometric properties of the network of mobile
particles when a space-dependent driving force (mimicking an applied
shear stress) is applied. This would be necessary to address the
transition from the shear-thinning to shear-thickening behavior in a
more realistic setup.  Second, the boundary conditions could be
modified by including static walls parallel to the transport
direction and different species of particles. That would allow one to
explore the effect of confinement and entropic sorting of
particles~\cite{entropicsorting}. This could be compared with the case
in which the wall is transversal to the applied field (as in granular
materials under gravity) where layering phenomena near the wall and
segregation effects have been observed~\cite{SeAr}.  Third, it would
be interesting to extend our approach to the case in which transport
is induced by external particle reservoirs~\cite{Se}.  Finally, one could modify our model by imposing velocity kinks randomly in space and time to the particles, the dynamics without the kinks obeying the same dynamical constraints as before: this would allow one to investigate the possible
emergence of congested traffic motion in active fluids~\cite{cugliandolo,wolynes,vicsek,
  active-part-review,giomi-liverpool-marchetti,heidenreich}.

\textit{Acknowledgements - }We are grateful to I. Neri for the
interpretation of equation (2), and to V. Lecomte, F. van Wijland,
J. Kurchan and L. Berthier for interesting discussions. FT is supported
by the French Ministry of Research and EP by CNRS and PHC No. 19404QJ.


\begin{thebibliography}{99}


\bibitem{DH} For reviews see: H. Sillescu, J. Non-Cryst. Solids {\bf
  243}, 81 (1999); M.D. Ediger, Annu. Rev. Phys.  Chem. {\bf 51}, 99
  (2000); S.C. Glotzer, J. Non-Cryst.  Solids, {\bf 274}, 342 (2000);
  R. Richert, J. Phys.  Condens. Matter {\bf 14}, R703 (2002);
  H. C. Andersen, Proc. Natl. Acad. Sci. U. S. A. {\bf 102}, 6686
  (2005).

\bibitem{heterogeneities}{\it Dynamical heterogeneities in glasses, colloids, and granular media}, edited by  L. Berthier, G. Biroli, J-P Bouchaud, L. Cipelletti and W. van Saarloos  (Oxford University Press, Oxford, 2011).

\bibitem{larson} R.G. Larson, {\it The Structure and Rheology of
  Complex Fluids} (Oxford University Press, Oxford, 1999).

\bibitem{fragile} {\em Soft and Fragile Matter: Non-equilibrium
  Dynamics, Metastability and Flow}, edited by M.E. Cates and
  M.R. Evans (IoP, London 2000).

\bibitem{jamming} {\em Jamming and Rheology: Constrained Dynamics on
  Microscopic and Macroscopic Scales}, edited by A. Liu and S.R. Nagel
  (Taylor \& Francis, London 2001).

\bibitem{Ritort-Sollich} F. Ritort and P. Sollich, Adv. Phys.  {\bf
  52}, 219 (2003).

\bibitem{kcms} J.P. Garrahan, P. Sollich and C. Toninelli, in \textit{Dynamical Heterogeneities in Glasses, Colloids, and Granular Media}, edited by L. Berthier, G. Biroli, J.-P. Bouchaud, L. Cipellietti, and W. van Saarloos (Oxford University Press Oxford, 2011), Chap. 10.
%
%
%
%
%
%



\bibitem{sellitto} M. Sellitto, Phys. Rev. Lett. {\bf 101}, 048301
  (2008). 

\bibitem{sellitto2} M. Sellitto, Phys. Rev. E {\bf 80}, 011134 (2009).
 
\bibitem{TurciPitard} F. Turci and E. Pitard, Europhys. Lett. {\bf 94},
  10003 (2011)
  
\bibitem{histories} J.P. Garrahan, R.L. Jack, V. Lecomte, E. Pitard,
  K. van Duijvendijk and F. van Wijland , Phys.Rev.Lett. {\bf 98}, (2007)
  195702. J. Phys. A {\bf 42}, (2009) 075007.  E. Pitard, V. Lecomte and
  F. Van Wijland, Europhys. Lett. {\bf 96}, 56002 (2011).
  

\bibitem{KoAn} W. Kob and H.C. Andersen, Phys. Rev. E {\bf 48}, 4364
  (1993).

\bibitem{KuPeSe} J. Kurchan, L. Peliti, and M. Sellitto,
  Europhys. Lett.  {\bf 39}, 365 (1997)

\bibitem{FrMuPa} S. Franz, R. Mulet and G. Parisi, Phys. Rev. E {\bf
  65}, 021506 (2002).

\bibitem{ToBiFi} C. Toninelli, G. Biroli, D.S. Fisher,
  Phys. Rev. Lett {\bf 92}, 185504 (2004).

\bibitem{MaPi} E. Marinari and E. Pitard, Europhys. Lett. {\bf 69},
  235 (2005).

\bibitem{ChSaKo} P. Chaudhuri, S. Sastry and W. Kob,
  Phys. Rev. Lett. {\bf 101}, 190601 (2008)

\bibitem{PaPiCaCo} R. Pastore, M. P. Ciamarra, A. de Candia and
  A. Coniglio, Phys. Rev. Lett. {\bf 107}, 065703 (2011)

\bibitem{asep} H. Spohn, J. Phys. A {\bf 16} 4275 (1983).


\bibitem{JKGC} R. L. Jack, D. Kelsey, J.P. Garrahan, D. Chandler,
  Phys. Rev. E {\bf 78}, 011506 (2008).

\bibitem{shearbands} F. Varnik, L. Bocquet, J. L. Barrat, L. Berthier,
  Phys. Rev. Lett. {\bf 90} , 095702 (2003).

\bibitem{Chaudhuri} P. Chaudhuri, L. Berthier, and W. Kob,
  Phys. Rev. Lett. {\bf 99}, 060604 (2007).

\bibitem{Stariolo} D.A. Stariolo and G. Fabricius, 
J. Chem. Phys. {\bf 125}, 64505 (2006).

\bibitem{shlovskii} E.I. Levin, B. I. Shklovskii, Solid State
  Communications, {\bf 67} (3), (1988).

\bibitem{bouchaud-anom-diff} J.-P. Bouchaud, in {\it Anomalous
  Transport: Foundations and Applications}, edited by R. Klages,
  G. Radons, and I.M. Solokov, Wiley-VCH (Berlin, 2008).
\bibitem {bouchaud-traps} C. Monthus, J.-P. Bouchaud, J. Phys. A {\bf
  29}, 3847 (1996).

\bibitem {Dhar-Dietrich} D. Dhar, D. Stauffer, Int. J.  Mod. Phys. C
  {\bf 9}, 349 (1998).


\bibitem{Jung:2004vo} Y. J. Jung, J.P. Garrahan  and D. Chandler
Phys. Rev. E.{\bf 69}, 061205 (2004).


\bibitem{Shokef} Y. Shokef and A. Liu,
Europhys. Lett. {\bf 90}, 26005 (2010).

\bibitem{Redner} A. Gabel, P.L. Krapivsky and S. Redner,
Phys. Rev. Lett. {\bf 105}, 210603 (2010).


\bibitem{entropicsorting} D. Reguera, A. Luque, P. S. Burada,
  G. Schmid, J. M. Rubi, and P. Hanggi Phys. Rev. Lett. {\bf 108},
  020604 (2012).


\bibitem{SeAr} M. Sellitto, and J.J. Arenzon, Phys. Rev. E {\bf 62},
  7793 (2000). Y. Levin, J.J. Arenzon and M. Sellitto,
  Europhys. Lett. {\bf 55}, 767 (2001).

\bibitem{Se} M. Sellitto,
Phys. Rev. E {\bf 65} 020101 (2002).

\bibitem{cugliandolo}	D. Loi, S. Mossa, and L. F. Cugliandolo, Physical Review E {\bf 77}, 051111 (2008).
\bibitem{vicsek} A. Czirok and T. Vicsek,  Physica A {\bf 281}, 17 (2000).


\bibitem{active-part-review} P. Romanczuk, M. Baer, W. Ebeling,
  B. Lindner, and L. Schimansky-Geier, Eur Phys J-Spec Top {\bf 202},
  1 (2012).

\bibitem{wolynes}S. Wang and P. G. Wolynes, Proc Natl Acad Sci USA {\bf 108}, 15184 (2011).
\bibitem{giomi-liverpool-marchetti}L. Giomi, T. B. Liverpool, and
  M. C. Marchetti, Phys. Rev. E {\bf 81}, 051908 (2010).

\bibitem{heidenreich} S. Heidenreich, S. Hess, and S. H. L. Klapp,
  Phys. Rev. E {\bf 83},011907 (2011).




\end{thebibliography}

\end{document}